\documentclass[%
 reprint,
superscriptaddress,
a4paper,
 amsmath,amssymb,
 aps,
 prl,
floatfix,
]{revtex4-2}

\usepackage{graphicx}
\usepackage{dcolumn}
\usepackage{bm}
\usepackage{siunitx}
\usepackage{multirow}
\usepackage[table,xcdraw]{xcolor}
\usepackage{xcolor}
\usepackage[english]{babel}

\begin{document}

\preprint{APS/123-QED}

\title{Unconventional bulk superconductivity in YFe$_2$Ge$_2$ single crystals}

\author{Jiasheng Chen}
\affiliation{Cavendish Laboratory, University of Cambridge, Cambridge CB3 0HE, United Kingdom}
\author{Monika B. Gam\.za}
\affiliation{Jeremiah Horrocks Institute for Mathematics, Physics and Astronomy, University of Central Lancashire, Preston, United Kingdom}
\author{Jacintha Banda}
\affiliation{Max Planck Institute for Chemical Physics of Solids, N\"othnitzer Stra\ss e 40, 01187 Dresden, Germany}
\author{Keiron Murphy}
\altaffiliation{Now at Clarendon Laboratory, Department of Physics, University of Oxford, Oxford, OX1 3PU, United Kingdom}
\affiliation{Cavendish Laboratory, University of Cambridge, Cambridge CB3 0HE, United Kingdom}
\author{James Tarrant}
\affiliation{Cavendish Laboratory, University of Cambridge, Cambridge CB3 0HE, United Kingdom}
\author{Manuel Brando}
\affiliation{Max Planck Institute for Chemical Physics of Solids, N\"othnitzer Stra\ss e 40, 01187 Dresden, Germany}
\author{F. Malte Grosche}
\affiliation{Cavendish Laboratory, University of Cambridge, Cambridge CB3 0HE, United Kingdom}

\date{\today}

\begin{abstract}
\noindent 
Using a new horizontal flux growth technique to produce high quality crystals of the unconventional superconductor YFe$_2$Ge$_2$ has led to a seven-fold reduction in disorder scattering, resulting in mm-sized crystals with residual resistivities $\simeq \SI{0.45}{\micro\ohm\cm}$, resistivity ratios $\simeq 430$ and sharp superconducting heat capacity anomalies. This enables searching multi-probe experiments investigating the normal and superconducting states of YFe$_2$Ge$_2$. Low temperature heat capacity measurements suggest a significant residual Sommerfeld coefficient, consistent with in-gap states induced by residual disorder as predicted for a sign-changing order parameter.
\end{abstract}

\maketitle


\noindent The iron-based layered compound YFe$_2$Ge$_2$ has been reported as a new unconventional superconductor \cite{Y_Zou2014}, which differs from other iron-based superconductors by its lack of group-V (pnictogen) or group-VI (chalcogen) elements and its more three dimensional Fermi surface geometry \cite{A_Subedi2014, D_J_Singh2014}. Nevertheless, YFe$_2$Ge$_2$ shares key  properties of the alkali metal iron arsenides, (K/Cs/Rb)Fe$_2$As$_2$: (i)~similarly enhanced Sommerfeld coefficient, $C/T \sim \SI{100}{\milli\joule/\mole~\kelvin^2}$ \cite{F_Hardy2014, A_F_Wang2013, S_Khim2017}; (ii)~bad metal behavior at room temperature, with resistivity $\rho$ of order $\SI{200}{\micro\ohm\centi\meter}$ \cite{Q_Si2016}; (iii)~strong suppression of the superconducting transition temperature $T_c$ by disorder scattering \cite{J-Ph_Reid2012}. The highly correlated low temperature state and its crossover to incoherent bad metal behavior at high temperature  \cite{F_Hardy2013, Y_P_Wu2016, D_Zhao2020} have been attributed to Hund's coupling, which forces the iron d-electrons into a high spin state, causing strong correlations and leading to narrow bands independently of the shape of the Fermi surface \cite{A_Georges2013}. 
Initially, no heat capacity anomalies were observed at the superconducting transition in polycrystals \cite{I_Felner2015} or single crystals \cite{H_Kim2015}, despite considerable efforts on optimizing the crystal growth. Further, systematic improvements of polycrystal quality produced the first thermodynamic evidence of bulk superconductivity in YFe$_2$Ge$_2$ \cite{J_Chen2016}, clarified the primary origin of crystalline imperfections in YFe$_2$Ge$_2$ and demonstrated that disorder rapidly suppresses the transition at a rate commensurate with expectations for a sign-changing order parameter \cite{J_Chen2019}. This motivated a renewed effort to produce bulk superconducting single crystals that are required for specialised experiments probing the strongly correlated normal state and the superconducting gap symmetry in YFe$_2$Ge$_2$. 

\begin{figure}[t]
\includegraphics[width=\columnwidth]{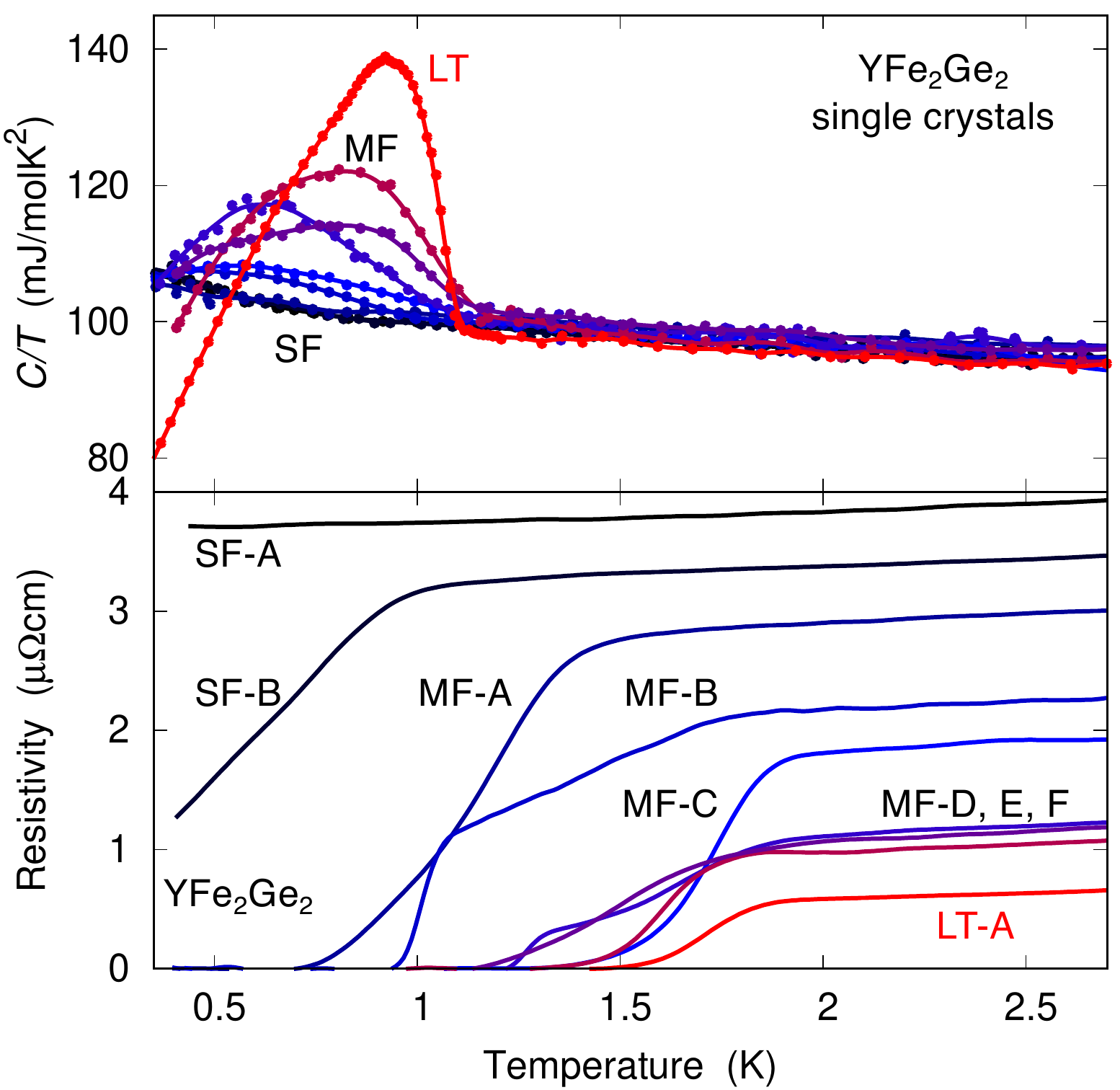}%
\caption{\label{FigureA} Sommerfeld ratio of the heat capacity $C/T$ (upper panel) and electrical resistivity (lower panel) of YFe$_2$Ge$_2$ single crystals from different growth batches. SF=standard flux growth, MF=modified flux growth, LT=liquid transport growth. These samples show a wide variation in their heat capacity superconducting anomaly, resistive transition and residual resistivity. Lower residual resistivities correlate with sharper superconducting heat capacity anomalies.}
\end{figure}

Here, we describe modified flux growth techniques which produce YFe$_2$Ge$_2$ single crystals of superior quality, with residual resistivity $\rho_0 < \SI{0.45}{\micro\ohm\centi\meter}$ and residual resistance ratio $\mathrm{RRR} \equiv \rho(\SI{300}{\kelvin})/\rho_0\sim 430$, establishing horizontal liquid transport as a powerful method for producing ultra-high quality crystals of challenging materials. Our crystals, for the first time, display sharp signatures of the superconducting transition in heat capacity, magnetisation and resistivity, proving that superconductivity is indeed intrinsic to YFe$_2$Ge$_2$ of sufficient quality. We find that the size of the superconducting heat capacity anomaly and the extrapolated residual Sommerfeld ratio $\gamma_0= \lim_{T\rightarrow 0} C/T$ correlate strongly with the disorder level and that $\gamma_0$ displays a further, distinct dependence on applied magnetic field. These results are consistent with unconventional superconductivity with a sign-changing gap function. 


{\em Methods --} Our crystal growth experiments are designed to enhance Fe occupancy on the Fe site, which in polycrystals proved to be the determining factor for achieving low disorder \cite{J_Chen2019}. A comprehensive crystal growth study using the conventional Sn-flux method demonstrated that lowering the growth temperatures leads to higher quality \cite{M_A_Avila2003, H_Kim2015}. This has motivated our approach, which reduces the temperature at which crystals precipitate from the melt by increasing the effective flux to charge ratio.
Multiple batches of YFe$_2$Ge$_2$ single crystals have been grown with Sn-flux methods using (i) standard flux growth (SF)  as described in \cite{P_C_Canfield1992}  and used in previous crystal growth studies on YFe$_2$Ge$_2$ \cite{M_A_Avila2003, H_Kim2015}, (ii) modified flux growth (MF) from a polycrystalline YFe$_2$Ge$_2$ precursor  
with a reduced peak temperature between 850 and $\SI{1150}{\degreeCelsius}$, or (iii) horizontal flux (or `liquid transport' - LT) growth across a temperature gradient in a two-zone furnace with the cold end at $\SI{500}{\degreeCelsius}$ 
\cite{SuppMat,J-Q_Yan2017}. The advantages of LT and MF growth vs. SF growth are discussed below. 
The in-plane electrical resistance was determined using a standard four-terminal {\em ac} technique in a Quantum Design Physical Properties Measurement System (PPMS) to $< \SI{0.4}{\kelvin}$. Data were scaled at \SI{300}{\kelvin} to the published high temperature resistivity of $\SI{190}{\micro\ohm\centi\meter}$ \cite{M_A_Avila2003}.
Heat capacity measurements employed the pulse-relaxation method in a PPMS to below \SI{0.4}{\kelvin} and a compensated heat pulse method in a $^3$He/$^4$He dilution fridge to $\SI{50}{\milli\kelvin}$ \cite{H_Wilhelm2004}. The magnetization data were taken in a Cryogenic SQUID magnetometer to below $\SI{0.45}{\kelvin}$ and were corrected for the effect of demagnetizing fields by approximating the sample shape as a rectangular prism.
Powder x-ray diffraction (XRD) patterns were collected at room temperature in the Bragg-Brentano geometry with Cu $K\alpha$ radiation at $\SI{40}{\kilo\volt}$ and $\SI{40}{\milli\ampere}$ on a Bruker D8 diffractometer equipped with a Lynxeye XE detector to reduce the effects of Fe fluorescence and $K\beta$ radiation. Refinements of the powder patterns 
showed no evidence of secondary phases except for occasional Sn flux inclusions. Lattice parameters were determined by referring to an internal Ge standard and using the Le Bail method.


\begin{figure}[t]
\includegraphics[width=0.9\columnwidth]{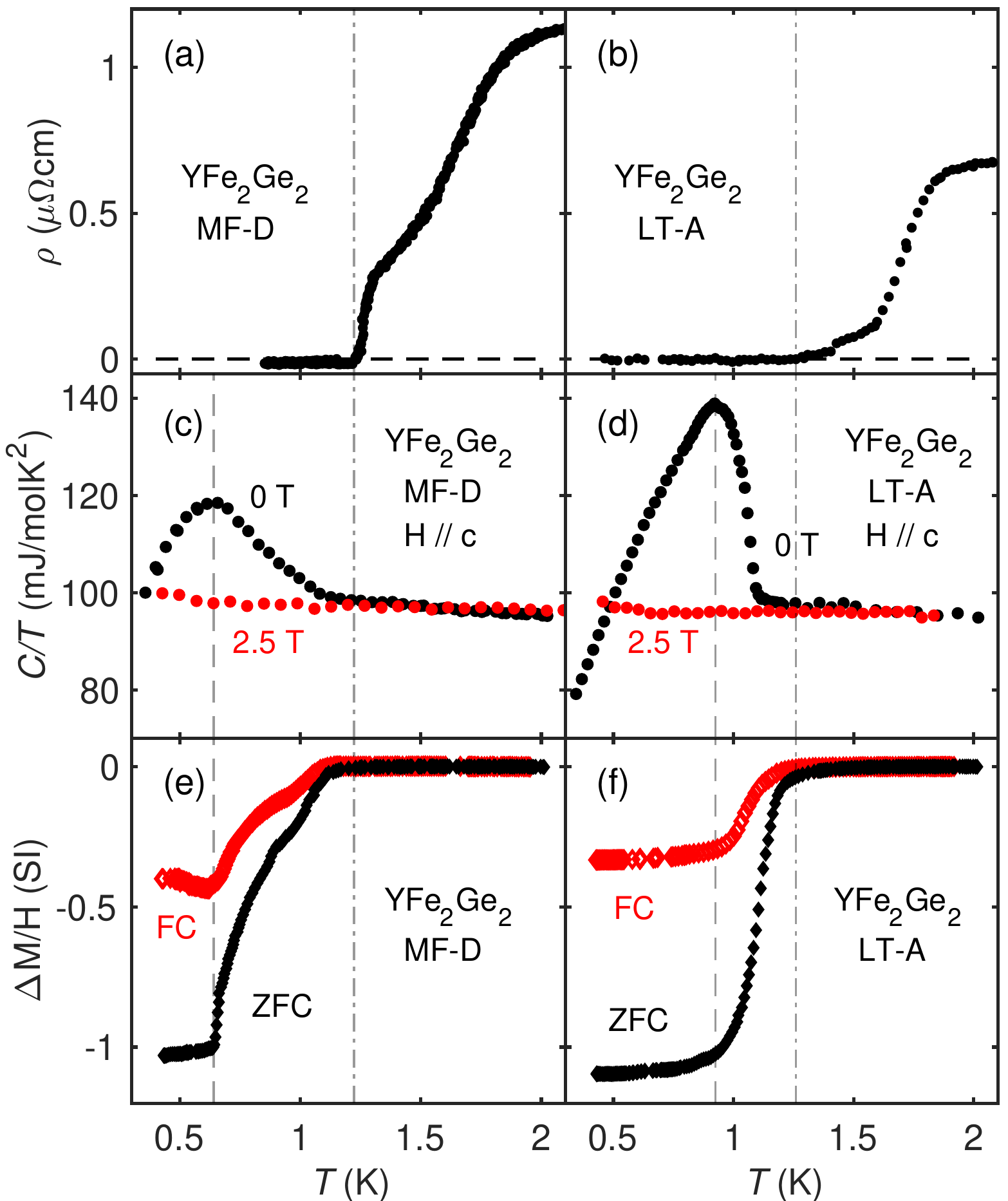}%
\caption{\label{FigureC} Electrical resistivity $\rho$ [(a) and (b)], Sommerfeld coefficient of the heat capacity $C/T$ [(c) and (d)] and magnetization (FC - field cooled, ZFC - zero-field cooled) [(e) and (f)] for two batches of high-quality YFe$_2$Ge$_2$ single crystals. The vertical dash-dotted lines denote the temperatures at which resistivity drops to zero in (a) and (b). The vertical dashed lines indicate the peaks of $C/T$ in (c) and (d).}
\end{figure}

\begin{figure}[t]
\includegraphics[width=\columnwidth]{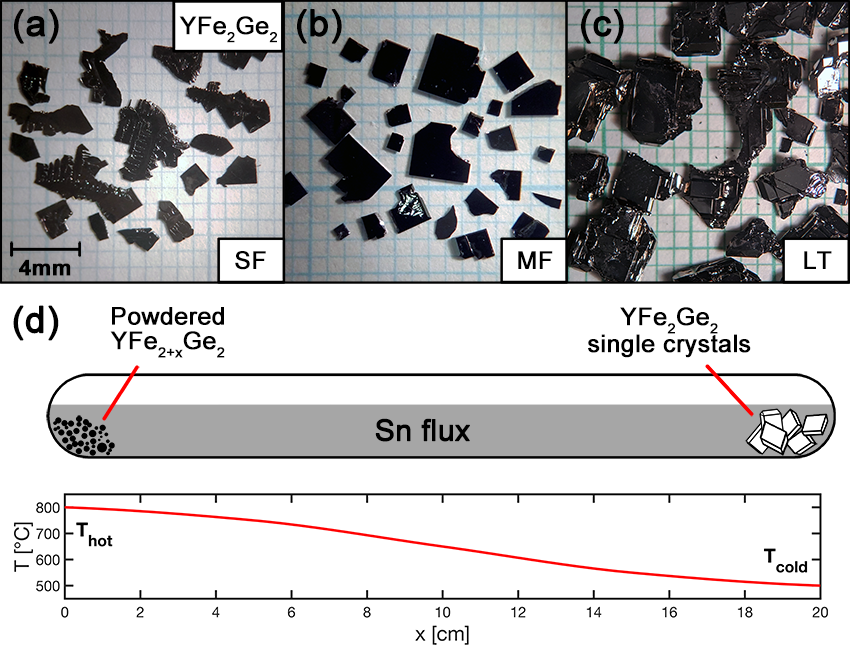}%
\caption{\label{FigureF} YFe$_2$Ge$_2$ single crystals produced using (a) SF, (b) MF and (c) LT methods on millimeter-scale paper. Dendrite formation is observed in the SF-grown crystals, which usually contain only one well defined facet along the [100] $a$-axis in each crystal. Crystals from the MF batches display less-obvious dendritic patterns, smoother surfaces and  more rectangular shapes. Both the SF- and MF-grown crystals form thin plates, whereas those from the LT batch are bulkier in the [001] c-direction with thickness reaching up to $\SI{2}{\milli\meter}$. (d) Schematic illustration of the liquid transport technique (LT).
}
\end{figure}

\noindent 


Low-temperature heat capacity and electrical resistivity of YFe$_2$Ge$_2$ crystals from different growth batches are compared in Fig.~\ref{FigureA}. Crystals grown using standard flux methods (SF) have extrapolated $T=0$ residual resistivity $\rho_0\simeq \SI{3.5}{\micro\ohm\cm}$, corresponding to $\text{RRR}\simeq 50-60$. Some of these samples show resistive transitions, but none display a superconducting heat capacity anomaly, confirming the findings of the previous single crystal study \cite{H_Kim2015}. The modified flux growth from polycrystal precursors (MF) leads to a marked improvement, reaching $\rho_0\simeq \SI{1}{\micro\ohm\cm}$ and $\text{RRR}\simeq 200$ in the best batches. These samples show full resistive transitions, which often split into two steps, and they show broad heat capacity anomalies similar to those observed in the best polycrystals with similar RRR values \cite{J_Chen2016, J_Chen2019}. Crystals grown in the liquid transport setup (LT) with a horizontal temperature gradient display the lowest residual resistivity $\rho_0\simeq \SI{0.45}{\micro\ohm\cm}$ and reach resistance ratios of up to $\text{RRR}\simeq 430$. These samples show sharp heat capacity jumps at a bulk $T_c$ of about $\SI{1.1}{\kelvin}$.

Resistive, thermodynamic and magnetic signatures of the superconducting transition in YFe$_2$Ge$_2$ in MF and LT samples are compared in Fig.~\ref{FigureC}.
An applied magnetic field of $\SI{2.5}{\tesla}$ completely suppresses superconductivity in both YFe$_2$Ge$_2$ samples (Figs.~\ref{FigureC}(c) and (d)), revealing the same underlying normal-state $C/T \approx \SI{100}{\milli\joule/\mol~\kelvin^2}$. SQUID magnetometry provides further evidence of bulk superconductivity in samples with $\text{RRR}\gtrsim 150$. 
The superconducting volume fraction inferred from the zero-field cooled diamagnetic response in Figs.~\ref{FigureC}(e) and~\ref{FigureC}(f) approaches $100\%$ just below the peak of the specific heat transition for samples from batch MF-D and LT-A respectively. 
The complete diamagnetic screening in the magnetisation measurement and the sharp transition in heat capacity both suggest 100\% bulk superconductivity in the high-quality samples. This is consistent with the mean free path $\ell$ and the scattering rate $\tau^{-1}$ extracted from resistivity measurements, using the approach described in \cite{J_Chen2019}: the residual resistivity $\rho_0 \simeq \SI{0.45}{\micro\ohm\centi\meter}$ of LT-grown crystals translates to $\ell \simeq \SI{2400}{\angstrom}$, far larger than the coherence length $\xi \simeq \SI{180}{\angstrom}$ \cite{J_Chen2016}, and to $\tau^{-1}\simeq \SI{0.09}{meV}/\hbar$, well below the threshold of about $\SI{0.2}{meV}$ expected within the Abrikosov-Gor’kov approach \cite{J_Chen2019,A_A_Abrikosov1961} from the bulk $T_c$.

\begin{figure}[t]
\includegraphics[width=\columnwidth]{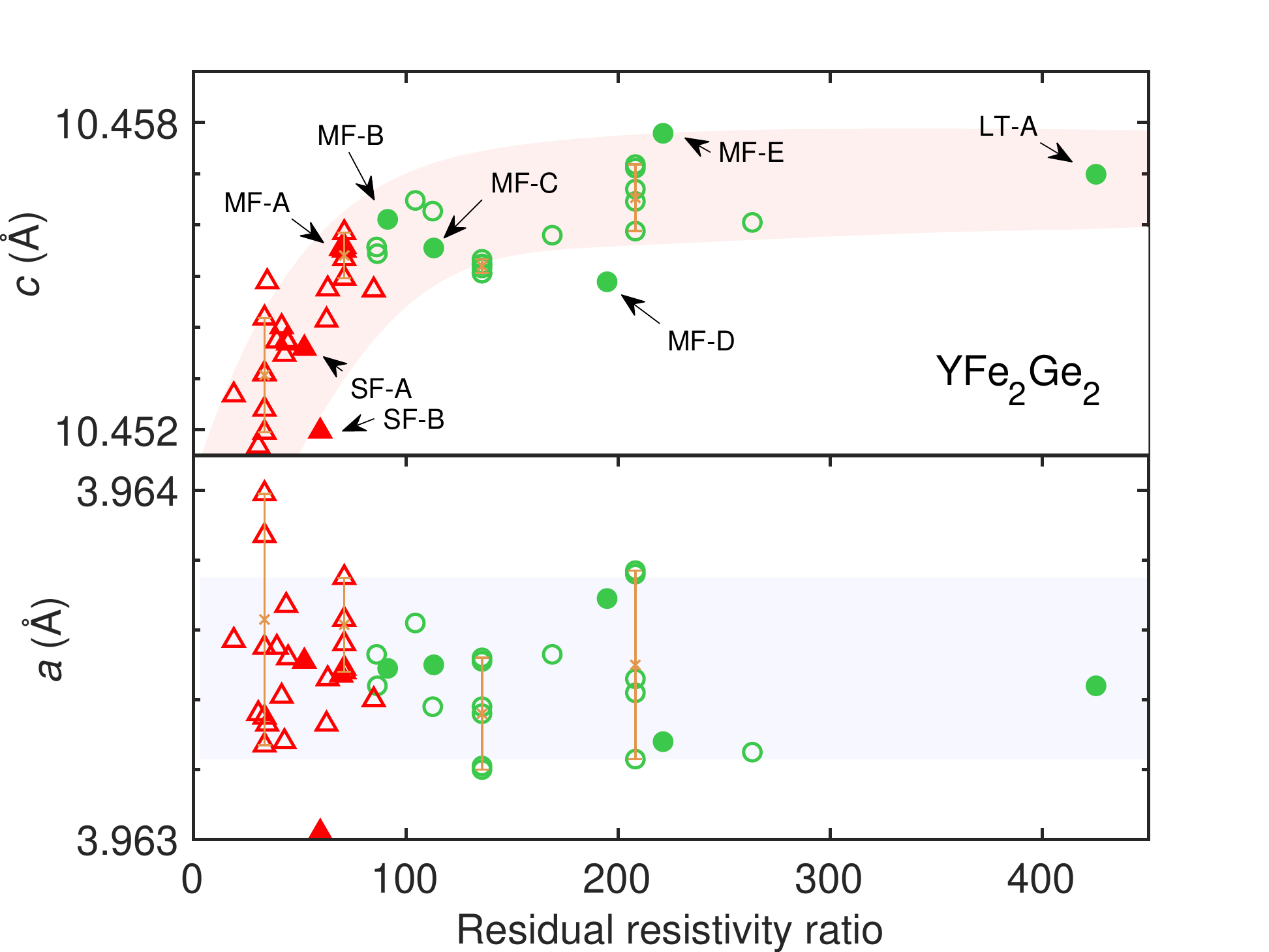}%
\caption{\label{FigureE} Lattice constants $c$ (upper panel) and $a$ (lower panel) of polycrystalline (open symbols)~\cite{J_Chen2019} and single-crystal (solid symbols) YFe$_2$Ge$_2$ samples, as obtained by powder XRD refinement, versus the corresponding RRR values. Green circles (red triangles) indicate that superconducting heat capacity anomalies have (have not) been observed. Errorbars are estimated from repeated measurements on selected batches of polycrystals. The pink and blue shades are guide to the eyes.}
\end{figure}



The three growth methods used in this study produce YFe$_2$Ge$_2$ crystals with different morphology, as illustrated in Fig.~\ref{FigureF}(a) to (c). Structural refinement of XRD data confirms an observation previously made in polycrystals \cite{J_Chen2019}, namely that 
improved crystal quality correlates with increasing c-axis lattice parameters (Fig.~\ref{FigureE}). This is consistent with Fe deficiency on the Fe site presenting the dominant source of disorder.
This interpretation is further supported by single-crystal XRD at $\SI{100}{\kelvin}$, which  
identifies significant Ge substitution on the Fe-site in the sample with the lowest RRR and c-axis lattice parameter (Batch SF-B)~\cite{SuppMat}, whereas bulk superconducting samples showed no signs of Ge-Fe site disorder. 

\begin{figure}[t]
\includegraphics[width=\columnwidth]{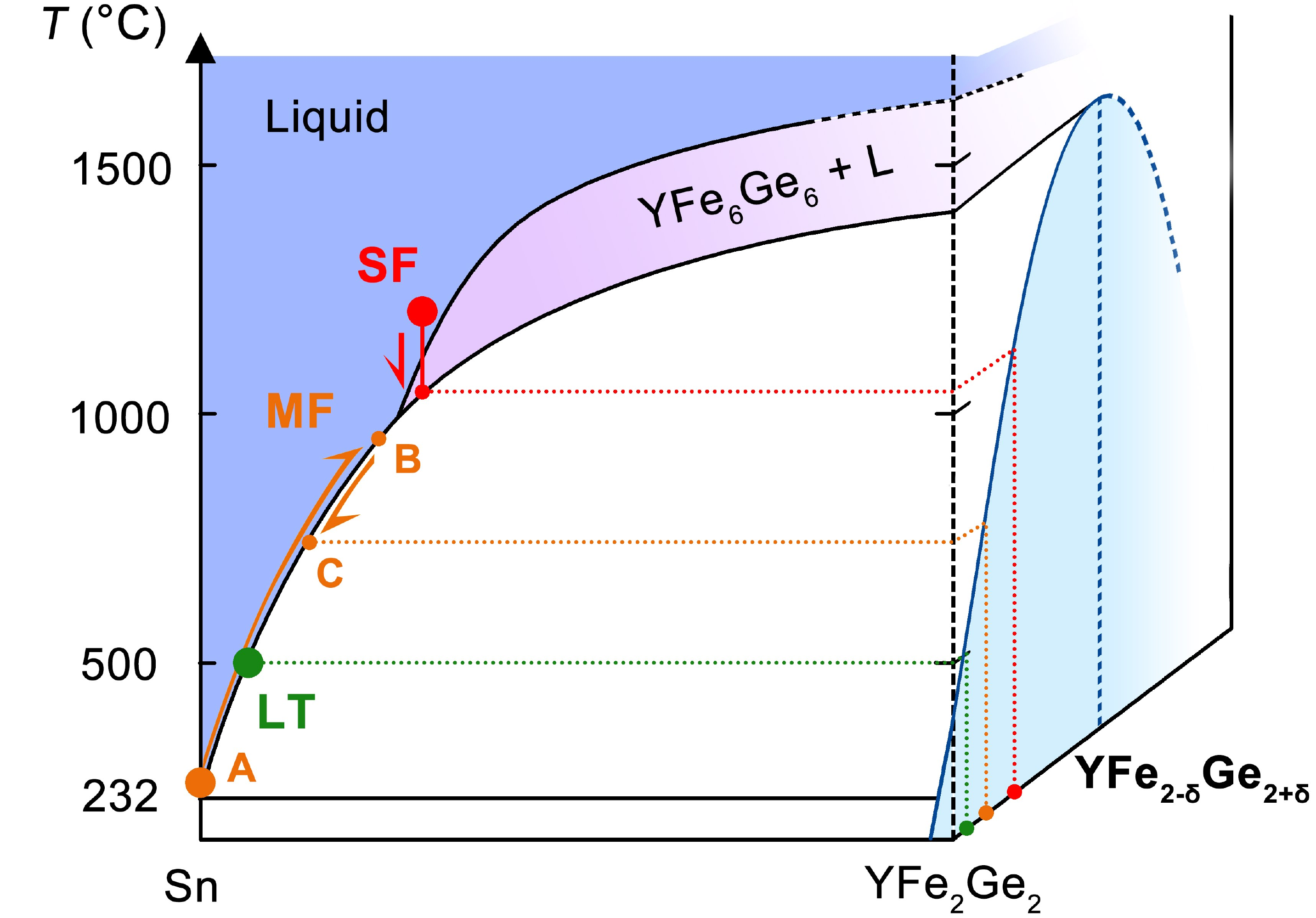}%
\caption{\label{FigureG} Schematic phase diagram illustrating YFe$_2$Ge$_2$ growth from Sn flux. SF growth proceeds from fully molten charge (red trajectory) and leads to Fe-poor crystals, if the apex of the YFe$_{2-x}$Ge$_{2+x}$ homogeneity range lies at an off-stoichiometric composition. MF growth proceeds from a saturated solution of YFe$_2$Ge$_2$ (orange trajectory), enabling a higher  flux/charge ratio in the melt and consequently lower precipitation temperatures. It produces crystals with closer to the ideal stoichiometry, but with considerable inhomogeneity. LT growth precipitates at a well-defined, low temperature and produces the highest quality crystals. At temperatures above $\sim \SI{1000}{\degreeCelsius}$ (purple zone), alien phases such as YFe$_6$Ge$_6$ may form, which affects the composition of the melt.}
\end{figure}

In standard flux growth, all elements are fully dissolved before crystals are formed on cooling. To achieve a high yield, this requires a low flux/charge ratio, which in turn leads to a high peak temperature, usually about $\SI{1200}{\degreeCelsius}$, and a high precipitation temperature. The schematic phase diagram in Fig.~\ref{FigureG} illustrates the benefits of our alternative growth protocols. 
We attribute the difficulties in growing high quality YFe$_2$Ge$_2$ crystals by the SF method to (i) the formation of secondary phases such as YFe$_6$Ge$_6$ at $T>\SI{1000}{\degreeCelsius}$, which affects the composition of the melt, and (ii) the position of the apex of the narrow YFe$_2$Ge$_2$ homogeneity range at a slightly off-stoichiometric composition YFe$_{2-\delta}$Ge$_{2+\delta}$, which favours Fe-poor crystals precipitating at high temperatures. 
In the modified flux method, the peak temperature can be reduced well below $\sim \SI{1000}{\degreeCelsius}$, while maintaining stoichiometry in the melt thanks to the polycrystalline precursor.
Because the precursor is not fully dissolved, the melt remains a saturated solution of YFe$_2$Ge$_2$ in Sn. The melt composition follows the liquidus curve from point A to point B as it heats up, gradually dissolving more of the charge. As the melt is cooled back down, it tracks the liquidus in reverse towards point C while precipitating YFe$_{2-\eta}$Ge$_{2+\eta}$ ($0 < \eta < \delta$). This process
limits the crystal growth to the desired low temperature region of the liquidus in the complex quaternary phase diagram, avoiding the formation of secondary phases and facilitating the growth of crystals with higher Fe/Ge ratio, closer to the stoichiometric composition. The heating/cooling cycle $C\rightarrow B\rightarrow C$ can be repeated to increase yield: every heating segment preferentially dissolves small crystallites on the polycrystalline charge, whereas each cooling segment preferentially deposits onto larger crystals forming elsewhere in the crucible. YFe$_2$Ge$_2$ is thereby gradually transported from the polycrystal precursor towards the single crystals, keeping the desired stoichiometry intact. Using Fe-rich precursors may further encourage high Fe content in the crystals.  However, because precipitation occurs  
over a range of temperatures, the resulting crystals exhibit some inhomogeneity of 
composition and disorder level, causing broad resistive and heat capacity anomalies. 
Liquid transport growth extends the idea of using the flux to transport material from precursor towards crystals to a horizontal geometry. 
Maintaining a constant temperature at the cold end of the flux mixture ensures that crystal growth takes place at a fixed point on the liquidus. This improves homogeneity and reduces Fe-Ge site disorder, resulting in the sharpened bulk transition in crystals from Batch LT-A.


\begin{figure}[t]
\includegraphics[width=0.9\columnwidth]{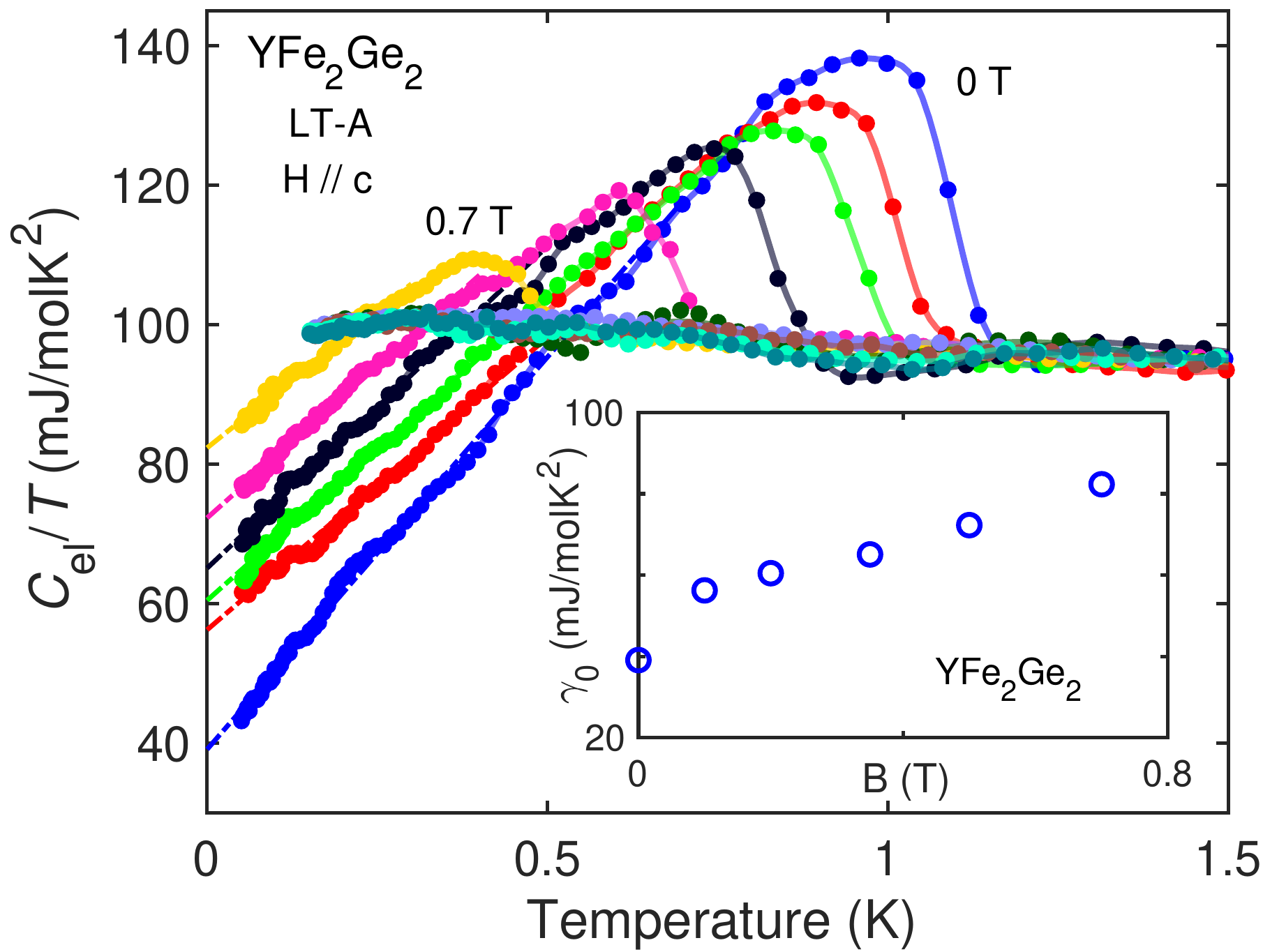}%
\caption{\label{FigureD} Electronic contribution to the specific heat capacity versus temperature, $C_{\text{el}}/T$, of a YFe$_2$Ge$_2$ single crystal from batch LT-A. Applied fields are 0 T (labelled), 0.1 T (red), 0.2 T (green), 0.35 T (black), 0.5 T (pink), 0.7 T (labelled) and 1 T (dark green). $C_{\text{el}}$ is obtained by subtracting from the total specific heat capacity the fitted nuclear contribution $C_\text{n}(H,T) = \alpha(H)/T^2$ \cite{SuppMat}. Inset: extrapolated zero-temperature residual $C_{\text{el}}/T$ versus applied field.}
\end{figure}

Low temperature heat capacity measurements (Fig.~\ref{FigureD}) in these samples
show a specific heat jump $\Delta C(T_c)/T_c$ of only $\sim 40\%$ of the normal state Sommerfeld coefficient $\gamma_n \simeq \SI{100}{\milli\joule/\mol~\K^2}$,  significantly less than $\Delta C(T_c)/T_c \approx 1.43 \gamma_n$ expected in a conventional $s$-wave superconductor but close 
to that observed in the isostructural compound KFe$_2$As$_2$ \cite{F_Hardy2014}. 
Extrapolating the linear temperature dependence of $C/T$ observed below $\SI{0.7}{\kelvin}$ to $T=0$ yields a residual Sommerfeld ratio $\gamma_0 \approx 0.4 \gamma_n$ at zero field, which rises further as the applied field is increased (inset to Fig.~\ref{FigureD}).

Our findings mirror key results in iron arsenides such as (Ba/K)Fe$_2$As$_2$. The fast suppression of bulk superconductivity by lattice disorder excludes a conventional isotropic $s$-wave scenario. The small $\Delta C(T_c)/\gamma_n T_c$ can be ascribed to multi-gap superconductivity, as in KFe$_2$As$_2$ \cite{K_Okazaki2012, F_Hardy2014}. Likewise, the significant extrapolated residual Sommerfeld ratio $\gamma_0$ might point towards a further downturn in $C_\text{el}/T$ at temperatures below the range of our current experiments, if the gap on part of the Fermi surface is much smaller than $k_B T_c$. However, because $\gamma_0$ rises with increasing disorder level, it is more plausibly attributed to impurity-induced states within the gap:
The magnitude of $\gamma_0/\gamma_n$ as well as its field dependence (Fig.~\ref{FigureD}) are similar to observations in some other iron-based superconductors (e.g. \cite{J_Kim2012}). For sign-changing order parameters, e.g. a two-band $s^{\pm}$-wave model, a finite $\gamma_0$ at zero field and a linear temperature dependence of $C/T$ at low $T$ -- as observed in YFe$_2$Ge$_2$  (Fig.~\ref{FigureD}) -- are expected when the level of impurity scattering increases above a critical threshold \cite{Y_Bang2017}. The field dependence of $\gamma_0$ is also consistent with this scenario \cite{Y_Bang2010,Y_Wang2011}, which is in line with earlier theoretical predictions \cite{A_Subedi2014} for YFe$_2$Ge$_2$.


Liquid transport growth emerges as a powerful tool for growing large, high quality crystals of YFe$_2$Ge$_2$ and other challenging materials. It has enabled a seven-fold reduction in disorder level in  the new generation of YFe$_2$Ge$_2$ crystals which, for the first time, show sharp thermodynamic transition anomalies indicative of bulk superconductivity. This will enable wide-ranging experiments probing superconducting and normal states in YFe$_2$Ge$_2$. The disorder sensitivity of the superconducting transition in general and of the residual Sommerfeld coefficient in particular point towards a sign-changing order parameter. 
 


\begin{acknowledgments}
We thank, in particular, A. Chubukov, C. Geibel, G. Lonzarich, J. Schmalian, F. Steglich and M. Sutherland for helpful discussions. The work was supported by the EPSRC of the UK (grants no. EP/K012894 and EP/P023290/1) and by Trinity College. 

\end{acknowledgments}

\bibliography{YFG2020Reference}

\end{document}


\onecolumngrid
\pagenumbering{arabic}
\renewcommand\thefigure{S\arabic{figure}} 
\setcounter{figure}{0} 
\renewcommand\thetable{S\arabic{table}} 
\setcounter{table}{0}

\begin{center}
{\bf{Supplementary Material for \\``Unconventional bulk superconductivity in YFe$_2$Ge$_2$ single crystals"}}\\
\vspace{4em}
\end{center}
\twocolumngrid

\noindent
This material details the crystal growth protocols used in growing different batches of YFe$_2$Ge$_2$ single crystals for which data is shown in the main manuscript and shows the single-crystal x-ray diffraction refinement results of samples from two growth batches. It also summarizes the method used for extracting the electronic contribution to the total specific heat capacity of YFe$_2$Ge$_2$ at low temperature.

\section{Flux growth of YF\MakeLowercase{e}$_2$G\MakeLowercase{e}$_2$}
\noindent
Table~\ref{TableS1} lists the starting compositions and temperature profiles for different YFe$_2$Ge$_2$ growth batches. In the standard flux (SF) growth, separate elements Y (4N; ALB Materials Inc.), Fe (4N, vacuum remelted; Alfa Aesar) and Ge (6N; Alfa Aesar) were used as the charge. The quartz ampoule, sealed with partial atmospheric pressure of argon, was heated to $\SI{1200}{\degreeCelsius}$ in order to completely dissolve the charge in Sn ($>$4N8; Alfa Aesar) flux. Starting atomic ratios significantly different from that used in \cite{H_Kim2015} have been adopted in the two SF batches with the initial goal to limit the growth of YFe$_6$Ge$_6$ secondary phase and to promote a high Fe occupancy in the YFe$_2$Ge$_2$ crystals. This has turned out to be ineffective in improving the crystal quality. In the modified flux (MF) and liquid transport (LT) growths, powderized polycrystal precursor produced in an induction furnace, with nominal composition YFe$_{2+x}$Ge$_2$ ($0.02 < x <0.1$), were instead used as the charge. The use of polycrystal precursors allows significant reduction in growth temperature in the MF growths. Comparison between peak growth temperatures and RRRs of crystals in different MF batches indicates a drastic improvement in sample quality with reduced peak temperature.

\section{Single-crystal X-ray diffraction}
\noindent
{\em Methods --} Single-crystal x-ray diffraction (XRD) was employed to investigate the structural differences between different batches of YFe$_2$Ge$_2$ crystals. Small irregularly shaped crystalline pieces suitable for the experiment were mechanically extracted from larger crystalline blocks. The diffraction measurements were performed at $\SI{100}{\kelvin}$ using a SuperNova Single Crystal Diffractometer (Rigaku Oxford Diffraction) with an Oxford Cryostream Cooler. Images acquired using an Atlas S2 $\SI{135}{\milli\meter}$ CCD Area Detector were processed using the CrysAlis$^{PRO}$ software. Details concerning data collection and handling are summarized in Table~\ref{TableS2}. Structure refinements were carried out using GSAS-II~\cite{B_H_Toby2013}.

Refinement of single-crystal XRD results confirmed the body-centred tetragonal ThCr$_2$Si$_2$-type structure (space group $I4/mmm$) of the YFe$_2$Ge$_2$ phase in crystals with all methods, in agreement with previous reports \cite{G_Venturini1996, H_Kim2015, J_Chen2019}. Details of structure refinements performed for crystals obtained from MF-D and SF-B batches are given in Table~\ref{TableS2}, and the resulting atomic positional, occupational and thermal displacement parameters are listed in Table~\ref{TableS3}.

For the crystal from Batch MF-D, the refinement indicated full occupancies of all the crystallographic sites and gave isotropic displacement parameters $B_{\mathrm{iso}}$(Y)~$\textless$~$B_{\mathrm{iso}}$(Ge)~$\textless$~$B_{\mathrm{iso}}$(Fe) (see Table~\ref{TableS3}), as expected based on the atomic masses. Attempts to further refine occupancies of all the lattice sites did not lead to any discernible improvement of the statistical factors. The occupation parameters of all sites converged to 100.0(2)\%, suggesting a stoichiometric composition of YFe$_2$Ge$_2$.

For the crystal from Batch SF-B, initial refinement performed assuming fully ordered crystal structure with complete occupancies of all crystallographic sites converged to a solution with low reliability factors $R$~=~3.41\%, $wR$~=~4.07\%. However, this refinement resulted in displacement parameters $B_{\mathrm{iso}}$(Ge)~$\textgreater$~$B_{\mathrm{iso}}$(Fe)~$\approx$~$B_{\mathrm{iso}}$(Y), which is unexpected, because Fe is lighter than Ge and Y. Furthermore, fast Fourier transformation revealed significant residual electron densities near 4d lattice sites. Therefore, the occupation parameter for the Fe position was further refined, which resulted in an occupation parameter of 101.5(2)\% and thus suggesting a partial replacement of Fe by heavier Y or Ge atoms. Since the unit cell volume of the SF-B crystal is smaller than that of the crystal from Batch MF-D, a mixed occupancy of the 4d lattice site by Fe and Ge is expected, as Ge atoms ($r$~=~1.22~\r{A}) are smaller compared to Fe ($r$~=~1.26~\r{A}) and Y ($r$~=~1.80~\r{A}) atoms \cite{J_Emsley1991_BOOK}. The refinement performed assuming that the 4d site is fully occupied by a random mixture of Fe and Ge atoms led to a notable lowering of the statistical factors to $R$~=~3.30\%, $wR$~=~4.01\% and gave thermal displacement parameters following the relation expected based on the atomic masses (see Table~\ref{TableS3}), with similar ratios of the displacement parameters to those obtained for the MF-D crystal. The resulting occupation parameters displayed in Table~\ref{TableS3} indicate that the chemical formula of the crystal from Batch SF-B is YFe$_{1.93(1)}$Ge$_{2.07(1)}$.

\section{Low-temperature specific heat}

\begin{figure}[t]
\includegraphics[width=0.9\columnwidth]{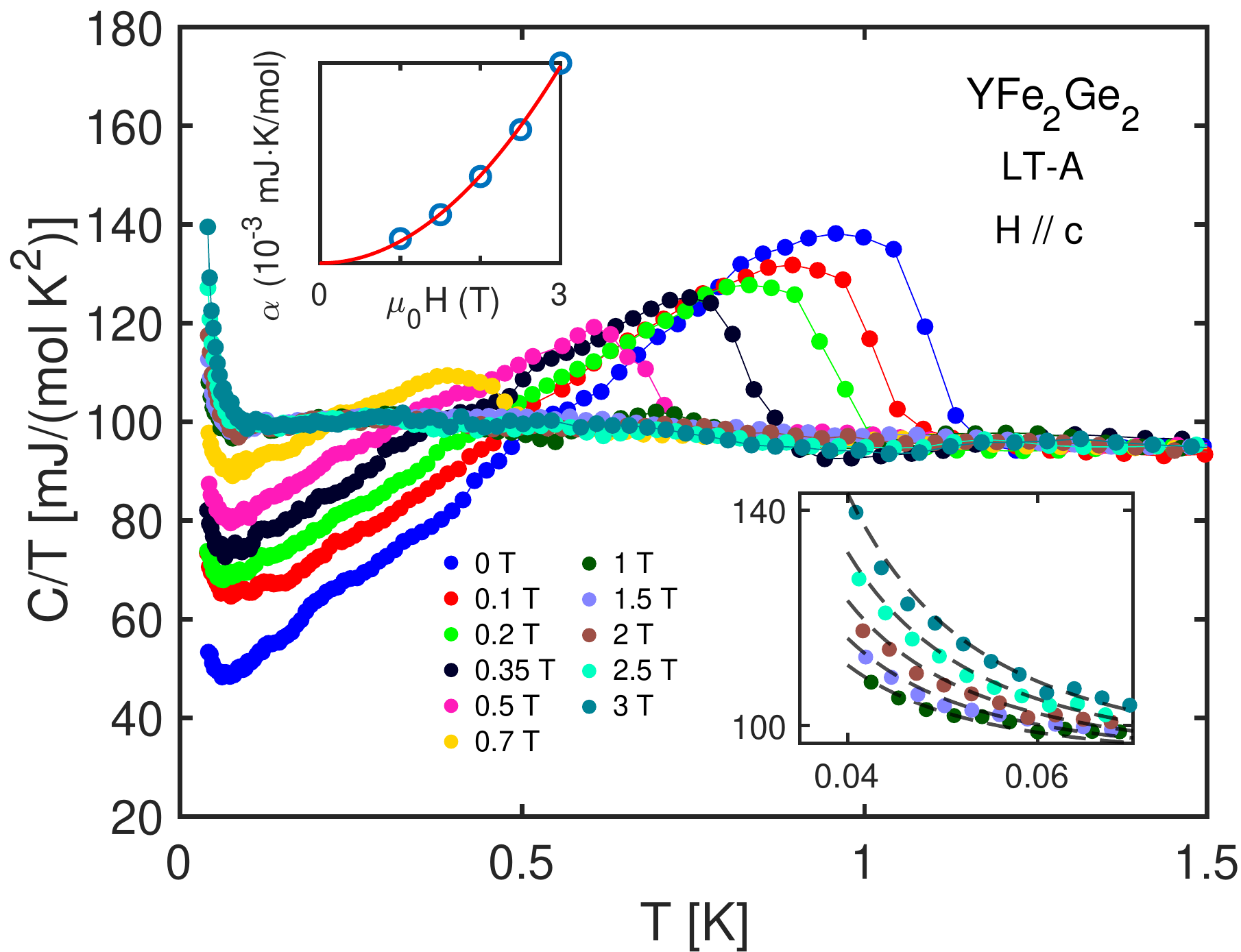}%
\caption{\label{FigureSA} Total heat capacity versus temperature $C/T$ of a YFe$_2$Ge$_2$ single crystal from batch LT-A. (Lower inset) $C/T$ at high fields of 1, 1.5, 2, 2.5 and $\SI{3}{\tesla}$ and the fits using the function $C/T = \alpha(H)/T^3 + \gamma $ (dashed lines).  (Upper inset) The field variation in the coefficient of the $T^{-3}$ term in the $C(T)/T$ data, $\alpha(H)$, as circles.}
\end{figure}

\noindent
Fig.~\ref{FigureSA} shows the total specific heat divided by temperature, $C/T$, of a YFe$_2$Ge$_2$ single crystal sample from LT-A batch, measured with various magnetic fields applied along the [001] $c$-axis. The total specific heat includes electronic, phonon, and nuclear contributions. To study the electronic part of the specific heat, subtraction of the other contributions is necessary. In this temperature range, the phonon contribution is negligible. On the contrary, the nuclear contribution, observed as an upturn in $C/T$ at low temperature, is significant at all fields. To evaluate the nuclear Schottky contribution, the field variation, $\alpha(H)$, in the coefficient of the $1/T^2$ term in $C(T)$ is found by fitting the normal-state data at 1, 1.5, 2, 2.5 and $\SI{3}{\tesla}$ between 40 and $\SI{65}{\milli\kelvin}$ using the function $C/T = C_n(H,T)/T + C_{\text{el}}/T = \alpha(H)/T^3 + \gamma$. The result is shown in the insets of Fig.~\ref{FigureSA} with $\alpha(H) \approx 0.88 + 0.26 H^2~ \si{\micro\joule\cdot\kelvin/\mole}$. The electronic contribution to the specific heat capacity shown in Fig.~5 of the main manuscript is obtained by subtracting $C_\text{n}$ from the total specific heat.

\begin{table*}
\caption{\label{TableS1}%
Atomic ratios of staring materials, temperature profiles used and the largest measured RRR values for all YFe$_2$Ge$_2$ growth batches discussed in the main manuscript. Batches are labelled SF/MF/LT (denoting standard flux method, modified flux method with polycrystalline charge and liquid transport method using horizontal configuration, respectively), followed by a letter.}
\renewcommand{\arraystretch}{2.3}
\begin{ruledtabular}
\begin{tabular}{cccc}
\multirow{2}{*}{Batch} & Starting ratio (at.)  & \multirow{2}{*}{Temperature profile} & RRR\\
 & [Y : Fe : Ge : Sn]  &  & [$\rho(\SI{300}{\kelvin})/\rho_0$]\\ \hline
SF-A & 1.4: 1.4 : 1 : 34 & \multirow{2}{*}{\begin{tabular}{c} \includegraphics[height = 0.8in]{GrowthProfiles/SF-A_TempProfile.pdf} \end{tabular}} & 52 \bigstrut \Tstrut\\
SF-B & 1.75 : 2.55 : 2 : 47  &  & 60\bigstrut \Bstrut\\ \hline
 & [YFe$_{2+x}$Ge$_2$\footnotemark[1] : Fe (powder): Sn] &  &  \TBstrut \\ \hline
MF-A & 1: 0.14 : 25  & \begin{tabular}{c} \includegraphics[height = 0.7in]{GrowthProfiles/MF-A_TempProfile.pdf} \end{tabular} & 70 \Tstrut\\
MF-B & 1: 0.1 : 29 \footnotemark[2] & \begin{tabular}{c} \includegraphics[height = 0.7in]{GrowthProfiles/MF-B_TempProfile.pdf} \end{tabular}  & 91\\
MF-C & 1: 0.38 : 39 & \begin{tabular}{c} \includegraphics[height = 0.7in]{GrowthProfiles/MF-C_TempProfile.pdf} \end{tabular}  & 113\\
MF-D & 1 : 0.19 : 31 & \begin{tabular}{c} \includegraphics[height = 0.7in]{GrowthProfiles/MF-D_TempProfile.pdf} \end{tabular} & 195\\
MF-E & 1 : 0.2 : 35 & \begin{tabular}{c} \includegraphics[height = 0.7in]{GrowthProfiles/MF-E_TempProfile.pdf} \end{tabular}  & 221\\
MF-F & 1 : 0.4 : 39  & \begin{tabular}{c} \includegraphics[height = 0.7in]{GrowthProfiles/MF-F_TempProfile.pdf} \end{tabular} & 207 \Bstrut\\ \hline
LT-A & 1 : 0.43 : 103  & \begin{tabular}{c} \includegraphics[height = 0.9in]{GrowthProfiles/LT-A_TempProfile.pdf} \end{tabular} & 425 \Tstrut
\end{tabular}
\end{ruledtabular}
\footnotetext[1]{Powdered YFe$_{2+x}$Ge$_2$ polycrystals ($0.02 < x < 0.1$) precursor produced using radio-frequency induction furnace.}
\footnotetext[2]{Additional yttrium pieces, with atomic ratio Y : YFe$_{2+x}$Ge$_2$ = 0.1 : 1, was added in the charge.}
\end{table*}

\begin{table*}
\caption{\label{TableS2} Single-crystal data collection and structure refinement parameters for crystals of the YFe$_2$Ge$_2$ phase from SF-B and MF-D batches.}
\setlength{\tabcolsep}{20pt}
\renewcommand{\arraystretch}{1.3}
\begin{tabular}{llll}
 \hline
 \hline
Refined stoichiometry  & YFe$_2$Ge$_2$ &  & YFe$_{1.93(1)}$Ge$_{2.07(1)}$ \\
Batch  & MF-D &  & SF-B \\
Structure type      & \multicolumn{3}{c}{ThCr$_2$Si$_2$} \\ 
Space group         & \multicolumn{3}{c}{$I4/mmm$ (No. 139)} \\
a (\r{A})  & 3.95917(3) & & 3.95873(5) \\
c (\r{A})  & 10.39754(13)  & & 10.39018(18)\\
V (\r{A}$^3$)  & 162.982(3) & & 162.830(4) \\
Crystal size ($\mu$m) &  106 $\times$ 39 $\times$ 28 & & 113 $\times$ 89 $\times$ 41 \\
Diffractometer & \multicolumn{3}{c}{SuperNova, precision four-circle Kappa} \\
                   & \multicolumn{3}{c}{Atlas S2 135mm CCD} \\
Radiation, $\lambda$ (\r{A}) & \multicolumn{3}{c}{Mo $K\alpha$, 0.71073} \\ 
Temperature (K) & \multicolumn{3}{c}{100(2)} \\ 
Range in $h,k,l$   & $\pm$7, $\pm$7, $\pm$20 & & $\pm$8, $\pm$8, $\pm$23 \\  
$R(eqv)$/$R(\sigma)$ & 0.0497/0.0494 & & 0.0518/0.0399 \\ 
2$\theta_{min}$/2$\theta_{max}$  & 3.92/45.53 & & 3.92/53.69 \\ 
Observation criteria & \multicolumn{3}{c}{$F$($hkl$) \textgreater 3.00$\sigma(F)$} \\
Absorption correction & \multicolumn{3}{c}{face-based, SCALE3 ABSPACK\cite{ABSPACK}} \\
Absorption coefficient & 44.311 & & 44.127 \\  
N($hkl$) measured & 25203  & & 13566 \\  
N($hkl$) unique    & 240  & & 159 \\
Extinction model & \multicolumn{3}{c}{Becker-Coppens secondary II Gaussian\cite{P_J_Becker1974}}  \\
Extinction coefficient (10$^{-6}$)  &  3.10(3) & & 5.10(6)\\
Number of parameters & 9 &  & 10 \\
Goodness-of-fit (GOF) & 1.32  &  & 1.71 \\
$R$                 & 3.26\%  &  & 3.30\% \\ 
$wR$                & 3.86\%  &  & 4.01\% \\ \hline \hline
 \end{tabular}
\end{table*}

\begin{table*}
\caption{\label{TableS3} Atomic positional, occupation and thermal displacement parameters for crystals of the YFe$_2$Ge$_2$ phase from MF-D and SF-B batches. (Note: $B_{12}=B_{13}=B_{23}=0$ and $B_{11}=B_{22}$ for all the atomic sites.)}
\setlength{\tabcolsep}{10pt}
\begin{tabular}{l|l|l|l|l|l|l|l|l}
\hline \hline
Atom &  Wyckoff site &  $x$   &  $y$   &  $z$        & $B_{11}$   & $B_{33}$ & $B_{\mathrm{iso}}$  &  Occupation \Tstrut \Bstrut\\ \hline  
      \multicolumn{9}{c}{MF-D batch: YFe$_2$Ge$_2$}  \Tstrut \Bstrut\\ \hline 
Y    &   2a  &   0    &   0    &   0         & 0.2301(22)& 0.3774(23) & 0.2932(17) & 1 \Tstrut\\
Fe   &   4d  &   0    &   0.5  & 0.25        & 0.2920(21)& 0.3442(24) & 0.3189(16) & 1 \\
Ge   &   4e  &   0    &   0    & 0.378335(7) & 0.2504(17)& 0.3656(22) & 0.3021(14) & 1 \Bstrut\\ \hline
      \multicolumn{9}{c}{SF-B batch: YFe$_{1.93(1)}$Ge$_{2.07(1)}$}  \Tstrut \Bstrut\\ \hline 
Y    &   2a  &   0    &   0    &   0         & 0.2116(47) & 0.3497(63) & 0.2629(39) & 1 \Tstrut\\
Fe/Ge&   4d  &   0    &   0.5  & 0.25        & 0.2724(63) & 0.3245(8)  & 0.2961(55) & 0.965(5)/0.035(5) \\
Ge   &   4e  &   0    &   0    & 0.378293(2) & 0.2266(34) & 0.3529(55) & 0.2755(39) & 1 \Bstrut\\ \hline
 \end{tabular}
\end{table*}

\section{Acknowledgments}

The authors thank Dr Nathan Halcovitch from Lancaster University for his helpful assistance during single-crystal XRD measurements.

\bibliography{YFG2020Reference}
\clearpage